\documentclass[letterpaper]{article}
\usepackage{natbib,alifeconf}
\usepackage{hyperref}
\usepackage{textgreek}
\usepackage{subfig}
\usepackage{bm}
\newcommand{\be}{\begin{eqnarray}}
\newcommand{\ee}{\end{eqnarray}}
\title{Flies as Ship Captains?\\ Digital Evolution Unravels Selective Pressures to Avoid Collision in Drosophila}
\author{Ali Tehrani-Saleh$^{1,2}$ \and Christoph Adami$^{2,3,4}$\\
\mbox{}\\
$^1$Computer Science and Engineering, Michigan State University  \\
$^2$BEACON Center for the Study of Evolution in Action  \\
$^3$Microbiology and Molecular Genetics, Michigan State University \\
$^4$Physics and Astronomy, Michigan State University, East Lansing, MI 48824 \\
adami@msu.edu}

\begin{document}
\maketitle

\begin{abstract}
Flies that walk in a covered planar arena on straight paths avoid colliding with each other, but which of the two flies stops is not random. High-throughput video observations, coupled with dedicated experiments with controlled robot flies have revealed that flies utilize the type of optic flow on their retina as a determinant of who should stop, a strategy also used by ship captains to determine which of two ships on a collision course should throw engines in reverse. We use digital evolution to test whether this strategy evolves when collision avoidance is the sole penalty. We find that the strategy does indeed evolve in a narrow range of cost/benefit ratios, for experiments in which the ``regressive motion" cue is error free. We speculate that these stringent conditions may not be sufficient to evolve the strategy in real flies, pointing perhaps to auxiliary costs and benefits not modeled in our study.
\end{abstract}

\section{Introduction}

How animals make decisions has always been an interesting, yet controversial, question to scientists~\citep{McFarland1977} and philosophers alike. Animals obtain various types of sensory information from the environment and then process these information streams so as to take actions that benefit them in survival and reproduction. Visual system plays an important role in providing animals with information about their environment, for example when foraging for food, detecting predators or prey, and when searching for potential mates. One of the primary components of visual information is motion detection. Motion is a fundamental perceptual dimension of visual systems~\citep{BorstEgelhaaf1989} and is a key component in decision making in most animals. Here, we study a very particular type of motion detection and concomitant behavior (collision avoidance) in {\it Drosophila melanogaster} (the common fruit fly), and attempt to unravel the selective (evolutionary) pressures that might have given rise to this behavior.

\begin{figure}
    \centering
    \includegraphics[scale=0.2]{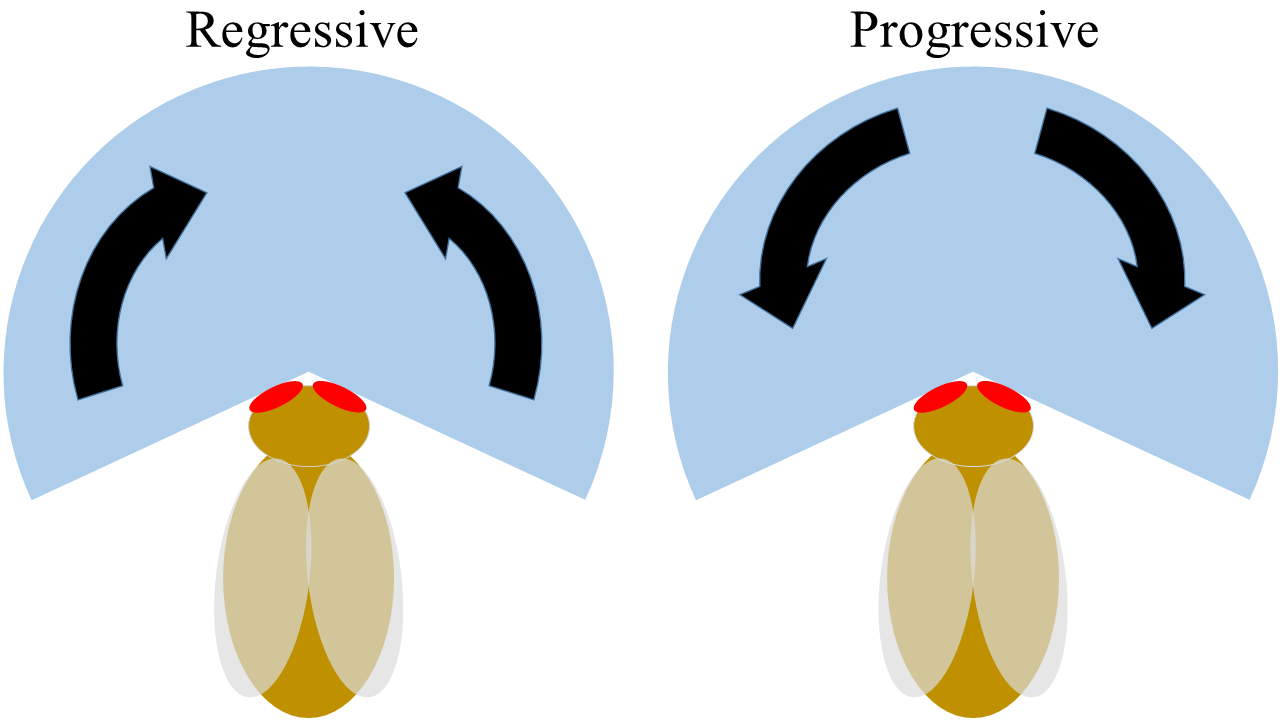}
    \caption{An illustration of regressive (back-to-front, left) and progressive (front-to-back, right) optic flows in a fly's retina.}
    \label{fig:regProg}
\end{figure}

 {\it D. melanogaster} shows a striking difference in behavior when exposed to two different types of optical flow. Branson and colleagues recorded the interaction of groups of fruit flies in a planar covered arena (so that they could only walk, not fly) and used machine-vision algorithms to analyze the walking trajectories in order to study fly behavior~\citep{branson2009high}. Their analysis revealed that female fruit flies stop walking when they perceive another fly's motion from back-to-front in their visual field (an optical flow referred to as ``regressive motion") whereas they keep walking when perceiving conspecifics moving from front-to-back in their visual field, referred to as ``progressive motion" (see figure~\ref{fig:regProg}). Zabala and colleagues further investigated this behavior and tested the ``regressive motion saliency" hypothesis, suggesting that flies stop walking when perceiving regressive motion~\citep{zabala2012simple}. They used a programmable fly-sized robot interacting with a real fly to exclude other sensory cues such as image expansion (``looming", see~\citealt{Schiffetal1962}) and pheromones. Their results provide rigorous support for the regressive motion saliency hypothesis. 
 
Subsequently, Chalupka and colleagues~(\citeyear{chalupka2015generalized}) coined the term ``generalized regressive motion" for optic flows in which images move clockwise on the left eye and conversely, counterclockwise on the right eye, see Fig~\ref{fig:regProg}. They presented a geometric analysis for two flies moving on straight, intersecting trajectories with constant velocities and showed that the fly that reaches the intersection first always perceives progressive motion on its retina, whereas the one that reaches the intersection later perceives regressive motion at all times before the other fly reaches the intersection. They went on to suggest that this behavior is a strategy to avoid collisions during locomotion similar to the rules that ship captains use when moving on intersecting paths (see, e.g.,~\citealt{Maloney1989}). 
As intriguing as this hypothesis may seem, it is not clear {\it a priori} which selective  pressures or environmental circumstances could give rise to this behavior. For example, it is unclear whether collision avoidance provides a significant enough fitness benefit. As a consequence, it is possible in principle that the behavior has its origin in a completely different cognitive constraint that is fundamentally unrelated to collision avoidance, or to the rules that ship captains use to navigate the seas. While such questions traditionally must be relegated to fire-side discussions amongst behavioral biologists, Artificial Life offers unique opportunities to test them directly.

In this study, we tested whether collision avoidance can be a sufficient selective pressure for the described behavior to evolve. We also investigated the environmental conditions under which this behavior could have evolved, in terms of the varying costs and benefits involved. By using an agent-based computational model (described in more detail below), we studied how the interplay (and trade-offs) between the necessity to move and the avoidance of collisions can result in the evolution of regressive motion saliency in digital flies.

Digital evolution is currently the only technique that can study hypotheses concerning the selective pressures necessary (or even sufficient) for the emergence of animal behaviors, as experimental evolution with animal lines of thousands of generations is infeasible. In digital evolution, we can study the interplay between multiple factors such as selective pressures, environmental conditions, population size and structure, etc. For example, Olson et al.~(\citeyear{olson2013predator}) used digital evolution to show that predator confusion is a sufficient condition to evolve swarming behavior, but they also found that collective vigilance can give rise to gregarious foraging behavior in group-living organisms~\citep{olson2015exploring}. In principle, any one hypothesis favoring the emergence of behavior can be tested in isolation, or in conjunction~\citep{olson2015exploring}.

\section{Methods}

\subsection{Markov Networks}

\begin{figure}
    \centering
    \includegraphics[width=\columnwidth]{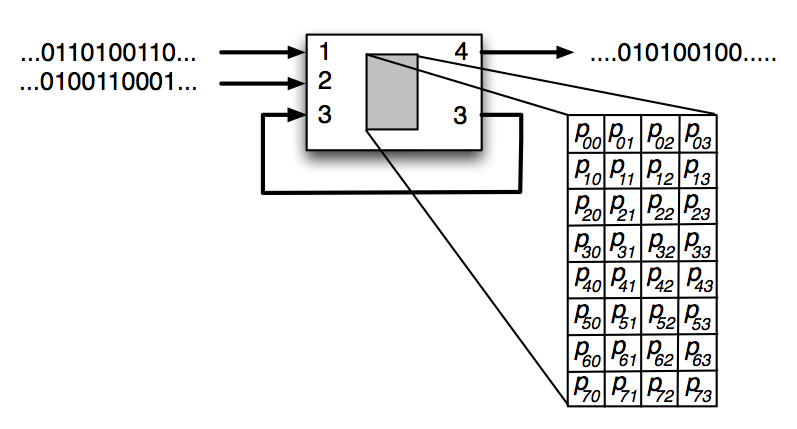}
    \caption{An illustration of probabilistic logic gates in Markov network brains. This gate has three inputs and two outputs. One of the outputs writes into one of the inputs of this gate, so its output is ``hidden". Because after firing all Markov neurons automatically return to the quiescent state, values can only be stored (kept in memory) by actively maintaining them. Probability table shows the probability of each output given input values.}
    \label{fig:plg}
\end{figure}

We use an agent-based model to simulate the interaction of walking flies with moving objects (here, potentially conspecifics) in a two-dimensional virtual world. Agents have sensors to perceive their surrounding world (details below) and have actuators that enable them to move in the environment. Agent brains in our experiment have altogether twelve sensors, three internal processing nodes, and one output node (the actuator). The brain controlling the agent is a ``Markov network brain (MNB)", which is a probabilistic controller that makes decisions based on sensory inputs and internal nodes~\citep{edlund2011integrated}  Each node in the network (i.e., sensors, internal nodes, and actuators) can be thought of as a digital (binary) neuron that either fires (value=1), or is quiescent (value=0). Nodes of the network are connected via hidden Markov gates (HMG) that function as probabilistic logic gates. Each HMG is specified by its inputs, outputs, and a state transition table that specifies the probability of each output state based on input states (figure \ref{fig:plg}). For example, in the transition table of figure~\ref{fig:plg} (a three-input, two-output gate), the probability $p_{73}$ controls the likelihood that the output state is 3 (the decimal equivalent of the binary pattern {\tt 11}, that is, both output neurons fire) given that the input happened to be state 7 (the decimal translation of {\tt 111}, i.e., all inputs are active). Markov brain networks can consist of any number of HMGs with any possible connection arrangement, given certain constrains (see for example~\citealt{edlund2011integrated}). 
\begin{figure*}[htbp!]
    \centering
    \includegraphics[width=\textwidth]{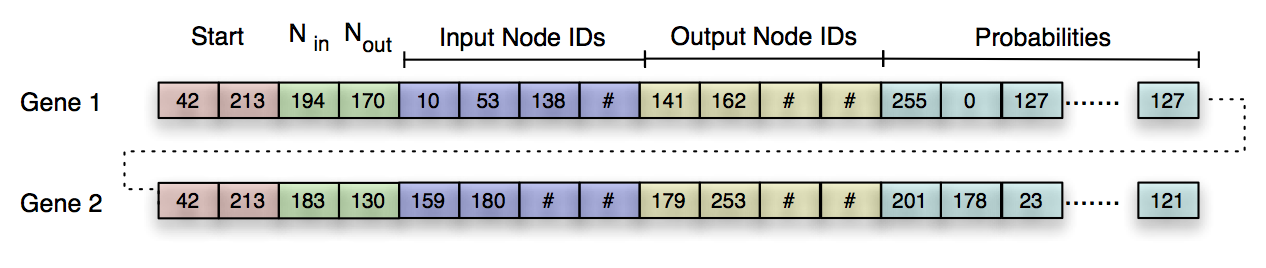}
    \caption{An illustration of a portion of genome containing two genes that encode two HMGs. The first two loci represent start codon (red blocks), followed by two loci that determine the number of inputs and outputs respectively (green blocks). The next four loci specify which nodes are inputs of this gate (blue blocks) and the following four specify output nodes (yellow blocks). The remaining loci encode the probabilities of HMG's logic table (cyan blocks).}
    \label{fig:encoding}
\end{figure*}

The number of gates, their connections, and how they work is subject to evolution and changes across individuals and through generations. For this purpose, the agent's brains are encoded in a genome, which is an ordered sequence of integers, each in the range [0,255], i.e. one byte. Each integer (or byte) is a locus in the genome and specific sequences of loci construct genes, where each gene codes for one gate. The ``start" codon for a gene (the sequence that determines the beginning of the gene) in our encoding is the pair (42,213) (these numbers are arbitrary). Each gene encodes exactly one HMG, for example as shown in figure~\ref{fig:encoding}. The gene specifies the number of inputs/outputs in each HMG, which nodes it reads from and writes to (the connectivity)  and the probability table that determines the gates' function. As shown in figure \ref{fig:encoding}, the first two bytes are the start codon, followed by one byte that specifies the number of inputs and one byte for the number of outputs. The bytes are modulated so as to encode the number of inputs and outputs unambiguously. For example, the bytes encoding the number of inputs is an integer in [0,255] whereas a Markov gate can take a maximum of four inputs, thus we use a mapping function that generates a number $\in$ [1,4] from the value of this byte. The next four bytes specify the inputs of the HMG, followed by another four bytes specifying where it writes to. The remaining bytes of the gene are mapped to construct the probabilistic logic gate table. The genomes are initialized with 5000 random integers and they contain four start codons to speed up the evolution at the beginning of the experiment. We can also force the gates to be deterministic rather than probabilistic (all values in the logic table are 0 or 1), which turns our HMGs into classical logic gates. Markov network brains have been used extensively in the last five years to study the evolution of navigation~\citep{edlund2011integrated,Joshietal2013}, the evolution of active categorical perception~\citep{Marstaller2013,albantakis2014evolution}, the evolution of swarming behavior as noted earlier, as well as how visual cortices~\citep{chapman2013evolution} and hierarchical groups~\citep{hintze2014evolution} form.

\subsection{Experimental Configurations}
We construct an initial population of 100 agents, i.e., digital flies, each with a genome initialized with 5,000 random integers containing four start codons (to accelerate early evolution). Agents (and by proxy the genomes that determine them) are scored based on how they perform in their living environment. The population of genomes is updated via a standard Genetic Algorithm (GA) for 50,000 generations, where the next generation of genomes is constructed via roulette wheel selection combined with mutations (detailed GA specifications are listed in table 1). There is no crossover or immigration in our GA implementation.

\begin{figure}
    \centering
    \includegraphics[width=\columnwidth]{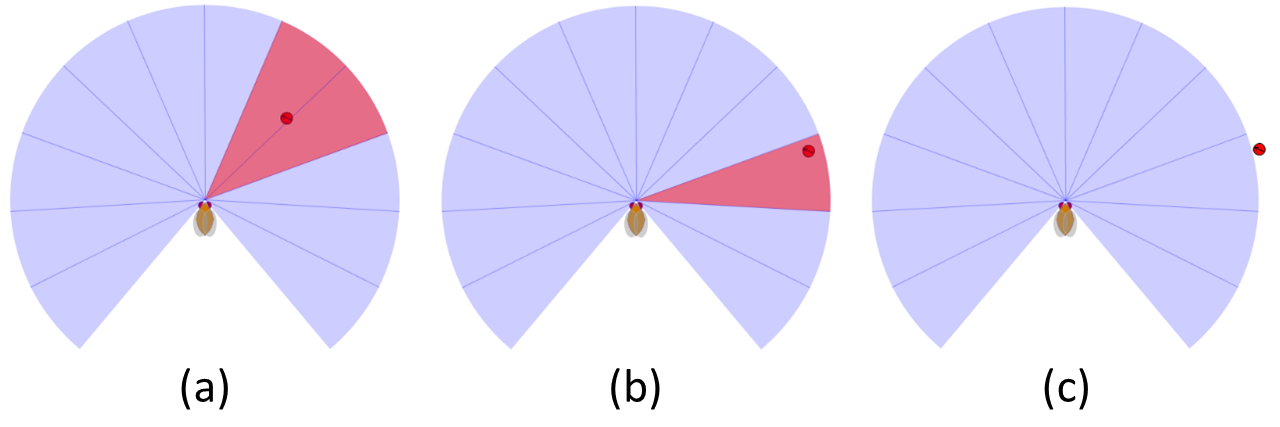}
    \caption{An illustration of a digital fly and its visual field in the model. Flies have a pixelated (12 pixels) retina that is able to sense surrounding objects in 280 degrees and within a limited distance (250 units). The red circle is an external object that can be detected by the agent within its vision field. Activated sensors are shown in red, while inactive sensors are blue. In (a) the object is outside of the visual field of the agent, in (b) the object is detected in one sensor, and in (c) the object activates two sensors.}
    \label{fig:retina}
\end{figure}

Each digital fly is put in a virtual world for 25,000 time steps where it can prove its prowess. During each time step in the simulation, the agent perceives its surrounding environment, and makes movement decisions. The sensory system of a digital fly is designed such that it can see surrounding objects within a limited distance of 250 units, in a 280 degrees pixelated retina shown in figure~\ref{fig:retina}. The state of each sensor node is 0 (inactive) when it does not sense anything within the radius, and turns to 1 (active) if an object is projected at that position in the retina. Agents in this experiment have one actuator node that enables them to move ahead or stop, for active (firing) and non-active (quiescent) states respectively.
\begin{table} [h]
\centering
\small
\begin{tabular}{|l|l|l|l|}
  \hline			
  \multicolumn{2}{|c|}{GA Parameters} &  \multicolumn{2}{|c|}{Environment Parameters}\\  
  \hline
   Population size   	& 100      &  Vision range &   250     \\
  \hline
   Generations          & 50,000  &  Field of vision    &   280 deg \\
  \hline            
   Point mutation rate& 0.5\% &  Collision range  & 60 \\
  \hline        
   Gene deletion rate   & 2\%  &  Agent velocity  & 15  \\
  \hline                                
   Gene duplication rate& 5\%  & Event time steps & 250   \\
  \hline
   Initial genome length& 5000 &  Number of events  & 100\  \\ 
  \hline  
   Initial start codons	& 4   &  Moving reward  & 0.004    \\
  \hline  
   Crossover         	& None  &   Collision penalty  & 1,2,3,5,10   \\
  \hline  
   Immigration			& None  & Replicates			& 20   \\
  \hline  
  \hline  
\end{tabular}
\caption{Configurations for GA and Environmental setup}
\end{table}
In our experiment, digital flies exist in an environment where they should move to gain fitness, representing the fact that organisms should forage for resources, mates, and avoiding predators. Thus, the fitness function is set so that agents are rewarded for moving ahead at each update of the world, and are penalized for colliding with objects. The amount of fitness they gain for moving (the benefit) is characteristic of the environment, and we change it in different treatments. The penalty for collisions represents the importance of collision avoidance for their survival and reproduction, and we vary this cost also. Each digital fly sees 100 moving objects (one at a time) during its lifetime, and we say that it experiences 100 ``events". The penalty-reward ratio (PR) determines the amount of penalty of collision divided by the reward for moving during the entirety of an event. So for example, PR=1 means the agent loses all the rewards it gained by walking during the whole event if it collides with the object in that event:
\be
\rm{fitness}=\sum_{\rm events} {\rm \left({reward} -PR\times \rm{collision}\right)} \;,
\ee
where $\rm{reward} \in [0,1]$ reflects how many time steps the agent moved during the event.
Our experiments are constructed such that {\em all} objects that produce regressive motion in the digital retina {\em will} collide with the fly if it keeps moving. The reason for biasing our experiments in this manner is explained in the following section.
\subsection{Collision Probability in Events with Regressive Optic Flow}

As mentioned earlier, Chalupka et al.~(\citeyear{chalupka2015generalized}) showed that for two flies moving on straight, intersecting trajectories with constant velocities, the fly that reaches the intersection first always perceives progressive motion on its retina while the counterpart that reaches the intersection later perceives regressive motion at all times before the first fly reaches the intersection. However, this does {\em not} imply that all objects that produce a regressive motion on a fly's retina will necessarily collide with it. In this section we present a mathematical analysis to discover how often objects that produce regressive motion in the fly's retina will eventually collide with the fly if the latter does not stop.

\begin{figure}
    \centering
  \subfloat{\includegraphics[width=.5\columnwidth]{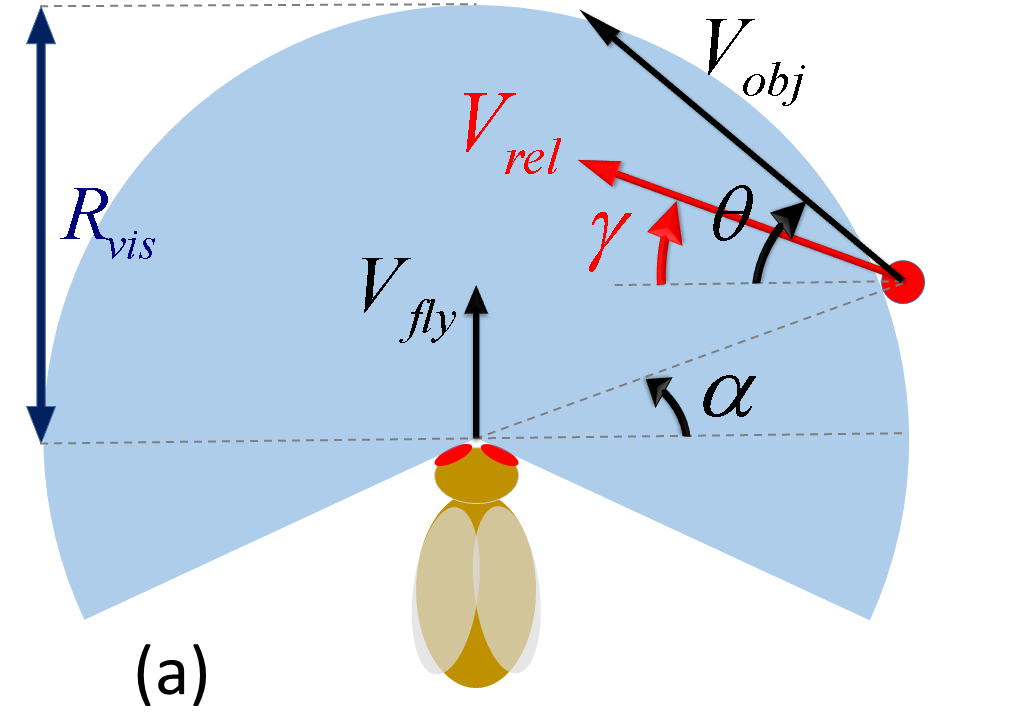}}
  \subfloat{\includegraphics[width=.5\columnwidth]{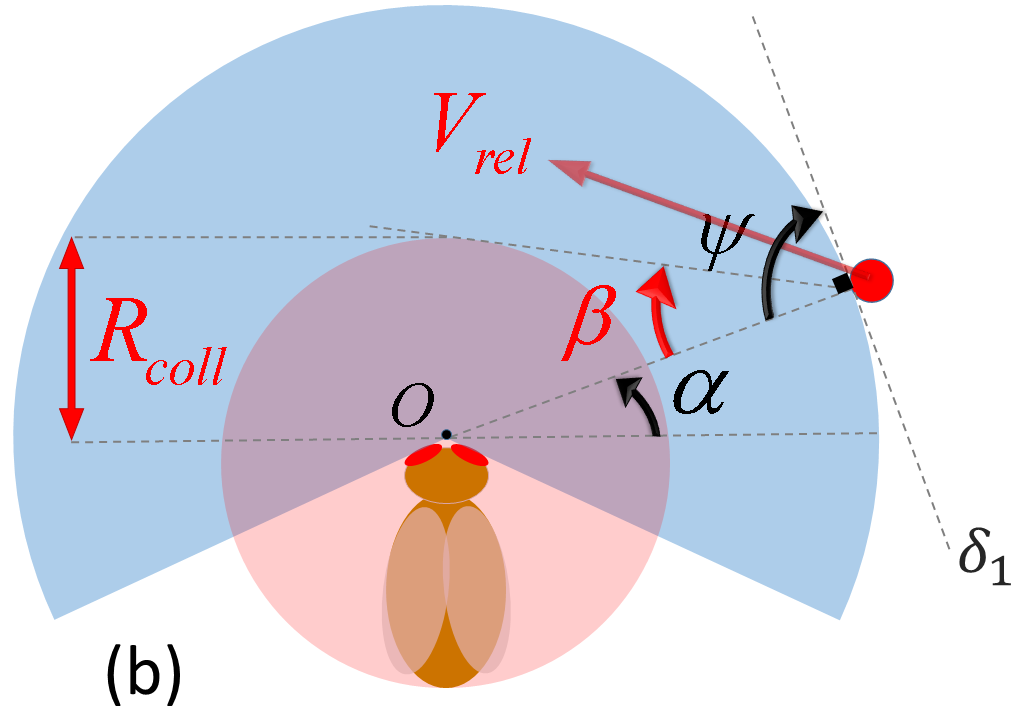}}
    \caption{An illustration of a moving fly at the onset of the event.}%
    \label{fig:col}%
\end{figure}

Suppose a fly moves on a straight line with constant velocity $\bm{V_{\rm fly}}$ and an object is also moving on a straight line with constant velocity $\bm{V_{\rm obj}}$ (figure \ref{fig:col}-a). The fly is able to perceive objects within distance $R_{\rm vis}$, its vision range (figure \ref{fig:col}-a). The object is assumed to be a point in the plane and the distance between this point and the center of the visual field of the fly is defined to be the distance between them. We define ``the onset of the event" as the first time the object is detected by the fly. At the onset of the event, the object is at the distance $R_{\rm vis}$ of the fly at relative azimuthal angle $\alpha \in [0,\frac{\pi}{2}]$ (figure~\ref{fig:col}-a). 
We assume that the object can be at any relative position $\bm{R_{\rm vis}}=(R_{\rm vis},\alpha$)\footnote{Here and below, we represent vectors either in boldface or by the parameters that determine them within a planar polar coordinate system. Thus the vector $\bf{R}$ is represented by $(|\bf{R}|,\phi)$, where $R_x=R\cos\phi$ and $R_y=R\sin\phi$.} with equal probabilities (the probability distribution of $\alpha$ is uniform around the fly). The velocity of the object can be represented as $\bm {V_{\rm obj}}=(V_{\rm obj} , \theta)$ where $\theta \in [\frac{-\pi}{2},\frac{\pi}{2}]$ (note that $\bm {V_{\rm obj}}$ is constant). We also assume that the velocity of the object can point in all directions with equal probabilities (the probability distribution of $\theta$ is uniform). The relative velocity of the object with respect to the fly is $\bm {V_{\rm rel}}=\bm{V_{\rm obj}} - \bm{V_{\rm fly}}$  (figure \ref{fig:col}). Since both $\bm{V_{\rm obj}}$ and $\bm{V_{\rm fly}}$ are constant, $\bm{V_{\rm rel}}$ is also a constant vector.
\subsubsection{Proposition 1.} A moving object produces regressive motion on a fly's retina if:
\be
\theta > -\alpha+\arcsin(\frac{V_{\rm fly}}{V_{\rm obj}} \cos\alpha)\;.
\ee
\subsubsection{Proof.} In order for the object to produce regressive motion on the retina, the relative velocity should be pointed above the center point O. The relative velocity direction $\gamma$ can be found awith $\bm{V_{\rm rel}}=(V_{\rm rel},\gamma)$, as
\be
\gamma=\arctan\left(\frac{V_{\rm rel_{\rm y}}} {V_{\rm rel_{\rm x}}}\right)=\arctan\left(\frac{V_{\rm obj}\sin \theta - V_{\rm fly}} {V_{\rm obj } \cos \theta}\right)\;.
\ee
The angle $\gamma$ should be greater than the central angle (figure \ref{fig:col}-b), that is, $\gamma > -\alpha$.
Replacing $\gamma$ and simplifying, we obtain:
\be
\theta > -\alpha+\arcsin(\nu \cos\alpha),\quad\nu=\frac{V_{\rm fly}}{V_{\rm obj}}\;.
\ee
For smaller values of  $\theta$, the object produces progressive optic flow. We thus define 
$\theta_{\rm min}=-\alpha+\arcsin(\nu \cos(\alpha))$ as the minimum angle $\theta$ that produces regressive motion on the retina.
\subsubsection{Definition 1.} The object remains ``observable" to the fly after the onset of the event if its relative velocity is directed toward the inside of the fly's vision field (to the left of the tangent line $\delta_1$ in figure \ref{fig:col}-b). 
\subsubsection{Proposition 2.} The object remains observable to the fly if:
\be
\theta<\arccos(-\frac{V_{\rm fly}}{V_{\rm obj}} \sin\alpha)-\alpha\;.
\ee
\subsubsection{Proof.} According to the definition the sufficient condition for observability is that $\gamma$ should be less than the tangent line $\delta_1$ angle:
$
\gamma < -\alpha+\frac{\pi}{2}\;.
$
Replacing $\gamma$ and simplifying we obtain
\be
\theta<\arccos(-\nu \sin\alpha)-\alpha\;.
\ee
For greater values of $\theta$, the object will be out of vision range of the fly. Thus the maximum value that $\theta$ can take on is:
\be
\theta_{\rm max}=\arccos(-\nu \sin\alpha)-\alpha\;.
\ee
In order for the object to produce regressive motion on fly's retina and also remain observable to the fly, relative velocity should be within the arc $\psi$ (figure \ref{fig:col}-b).
\subsubsection{Definition 2.}
The object collides with the fly if its distance with the fly is less than ``collision range" $R_{\rm coll}$ (figure \ref{fig:col}-b).
\subsubsection{Proposition 3.} An object that creates regressive optic flow on the fly's retina and remains observable will collide with it if:
\be
\theta < \phi + \arcsin(\nu \cos\phi), \quad \phi=\arcsin(\frac{R_{\rm coll}}{R_{\rm vis}})-\alpha\;.
\ee
\subsubsection{Proof.} The relative velocity of such object is within arc $\psi$. This object will collide with the fly if its relative velocity is in the range $\beta$, i.e. lower than tangent line to collision circle (figure \ref{fig:col}-b). This condition holds true if:
\be
\gamma < \beta - \alpha, \quad \beta=\arcsin(\frac{R_{\rm coll}}{R_{\rm vis}})\;.
 \ee
Let
$\rho=\frac{R_{\rm coll}}{R_{\rm vis}}$ and $\phi=\beta-\alpha$. 
Replacing $\gamma$ and rearranging gives:
\be
\theta < \phi + \arcsin(\nu \cos\phi)\;.
\ee
For greater values of $\theta$, the object produces regressive motion on the fly's retina but does not collide with it. So the threshold collision angle is given by:
\be
\theta_{\rm col}=\phi + \arcsin(\nu \cos\phi)\;.
\ee
As mentioned, we assume that the probability distribution of the direction of the object velocity, $\theta$ is uniform.
\subsubsection{Definition 3.} For an object at initial position $\alpha$, the probability $\Pi_{\rm coll}$ is the range of velocity directions $\theta$ such that the object collides with the fly divided by the range of directions with which it creates regressive optic flow on fly's retina (see figure \ref{fig:col}-b):
\be
\Pi_{\rm coll}(\alpha,\nu,\rho)=\frac{\theta_{\rm col}-\theta_{\rm min}}{\theta_{\rm max}-\theta_{\rm min}}\;.
\ee
Integrating this function over the range of possible initial relative positions, the probability that an event results in a collision given that the object produces regressive motion on an fly's retina can be found as:
\be
\Pi_{\rm coll}(\nu,\rho)=\int\limits_{\alpha_{\rm min}}^{\alpha_{\rm max}} \Pi_{\rm coll}(\alpha,\nu,\rho)d\alpha\;, \label{integral}
\ee
where $\alpha_{\rm min}$ is either 0 or the minimum value of $\alpha$ for which there exists a $\theta$ with which the object can produce a regressive motion on fly's retina, and $\alpha_{\rm min}$ is either 90 or maximum value of $\alpha$ for which there exists a $\theta$ with which the object remains observable to the fly.

\begin{figure}
    \centering
    \includegraphics[width=\columnwidth]{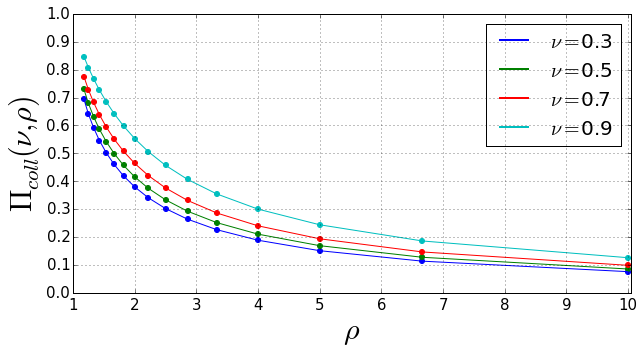}
    \caption{Probability of collision $\Pi_{\rm coll}(\nu,\rho)$ with an object that creates regressive motion on the retina as a function of the ratio of vision radius to collision radius $\rho$, for different fly-object velocity ratios $\nu$.}
    \label{fig:col-prob}
\end{figure}

We calculated the integral (\ref{integral}) numerically and show the results in figure \ref{fig:col-prob} for different values of fly-object velocity ratios $\nu$ and different collision range-vision range ratios $\rho$. As can be seen from figure \ref{fig:col-prob}, for $R_{\rm vis}$=60 mm~\citep{zabala2012simple} and $R_{\rm coll}$=15 mm (our assumption), the collision probability is around 0.2-0.3. This implies that if encounters are created randomly, regressive motion on the retina is not predictive of collision, and as a consequence it is unreasonable to expect that digital evolution will produce collision avoidance in response, as only 1 in 5 to 1 in 3 regressive motions actually lead to collisions. This was borne out in experiments, and we thus decided to bias the events in such a manner that {\em all} events that leave a regressive motion signature in the retina will lead to collision. Note that this is not necessarily an unrealistic assumption, as we have not analyzed a distribution of realistic ``events" (such as is available in the data set of \citep{branson2009high}). It could very well be that the way real flies approach each other differs from the uniform distributions that went into the mathematical analysis presented here.

\section{Results}
We conducted experiments with five different fitness functions representing different environments. Environments differ in the amount of fitness individuals gain when moving and in the penalty incurred by a collision. Evolved agents use various strategies to avoid collisions and maximize the travelled distance, but one of the most successful strategies they use is indeed to categorize visual cues into regressive and progressive optic flows. We find that agents categorize these visual cues only in some regions of the retina: the regions in which collisions take place more frequently. They then use this information to cast a movement decision: they keep moving when seeing an object creating progressive optic flow on their retina, and stop when the object creates regressive optic flow on their retina. However, they do not stop for the entire duration of the event, i.e., the whole time they perceive regressive optic flow. Rather they stop during only a portion of the event, which helps the agent to avoid a collision with the object while maximizing their walking duration and hence gaining higher fitness. 

The strategy of using regressive motion as a cue for collision~\citep{chalupka2015generalized}, similar to the observed behavior in fruit flies~\citep{zabala2012simple} thus evolves in our experimental setup under some environmental circumstances (discussed below). We refer to this strategy as regressive-collision-cue (RCC) and we define it in our experimental setup as follows:\\ 
\noindent 1) if the moving object produces regressive motion on agent's retina during an event and the agent stops at least for some amount of time during that event, or\\
 2) if the moving object produces progressive motion on agent's retina during an event and the agent does not stop at all during that event. The number of events (out of 100) in which the agent uses this strategy is termed the ``RCC value".
 
\begin{figure}
    \centering
    \includegraphics[width=\columnwidth]{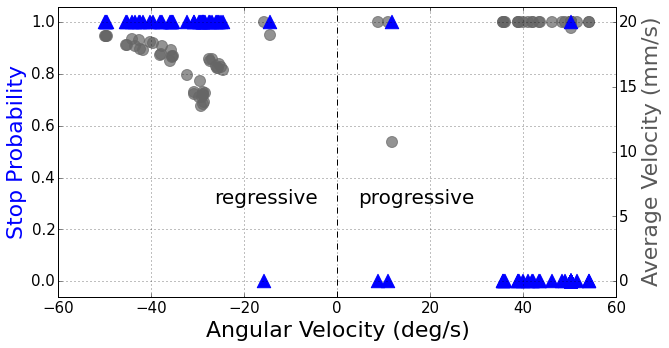}
    \caption{The stop probability of the evolved agent vs. the angular velocity of the image on its retina for 100 events. Positive values of angular velocity show progressive motion events and negative angular velocities stand for regressive motion events. The average velocity of the agent is also shown during each event.}
    \label{fig:stop-V}
\end{figure}

\begin{figure}
    \centering
    \includegraphics[width=\columnwidth]{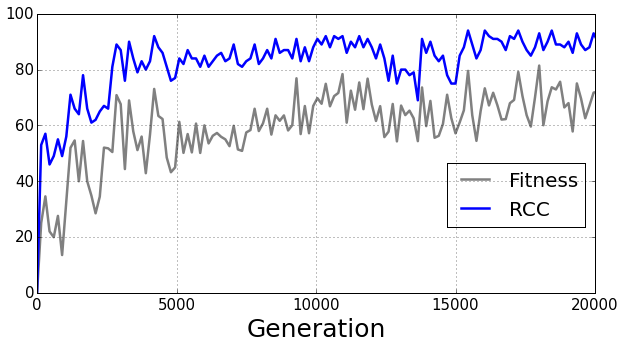}
    \caption{Fitness and regressive-collision-cue (RCC) value on the line of descent for an agent that evolved RCC as a strategy to avoid collisions. Only the first 20,000 generations are shown for every 150 generations.}
    \label{fig:lod_9}
\end{figure}

\begin{figure}
    \centering
    \includegraphics[width=\columnwidth]{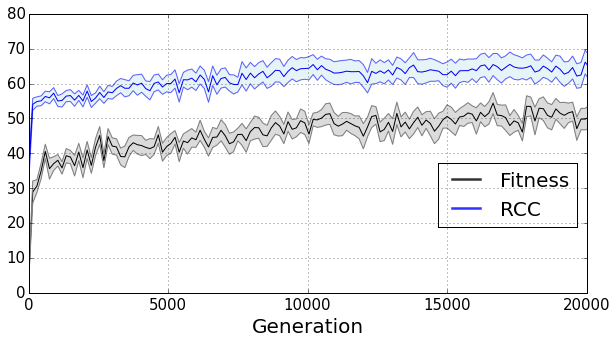}
    \caption{Mean values of fitness and regressive-collision-cue (RCC) over all 20 replicates vs evolutionary time in the line of descent in the environment with penalty-reward ratio of 2. Standard error lines are shown with shaded areas around mean values. Only the first 20,000 generations are shown for every 150 generations.}
    \label{fig:lod_p2}
\end{figure}

\begin{figure}
    \centering
    \includegraphics[width=\columnwidth]{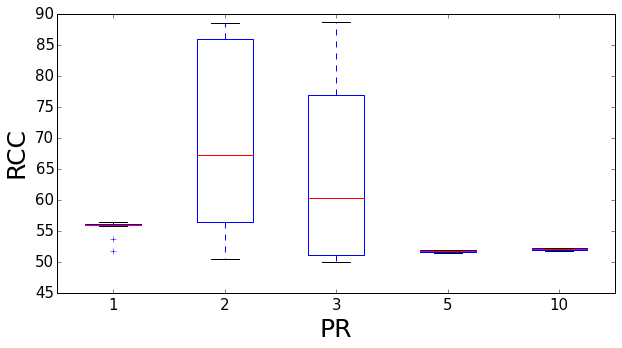}
    \caption{RCC value distribution in environments with different penalty-reward ratios. Each box-plot shows RCC value (averaged over last 1000 generations on the line descent) for 20 replicates.}
    \label{fig:rcc-env}
\end{figure}

We now discuss the results of an experiment in which the RCC strategy has evolved. We take the most successful agent at the end of that experiment and analyze its behavior. This agent evolved in an environment with penalty-reward ratio of 2, meaning the penalty of each collision equals twice the maximum reward the agent can gain in 2 events. Figure \ref{fig:stop-V} shows whether the agent stopped during an event, stop probability (in blue), as a function of the angular velocity of the image on the agent's retina for 100 events. In that figure, the angular velocity of the image on agent's retina is positive for regressive optic flow and negative for progressive events. Simulations units are converted to plotted values (in deg/s and mm/s) by equalizing dimensionless values $\nu$, and $\rho$ in simulation and actual values: $R_{\rm vis}$=60 mm \citep{zabala2012simple}, $V_{\rm fly}$=20 mm/s~\citep{zabala2012simple}, $R_{\rm coll}$=15 mm (our assumption). It can be seen from the figure that out of all 100 events, during one event with a regressive motion it did not stop at all and for two progressive events it stopped. In the remaining events it uses the RCC strategy (RCC value=97). The average velocity of the agent during each event is also shown (in grey), which reflects the number of time steps the agent moves during that event, and thus indirectly how often it stops. For progressive motions, the stop probability is zero (the agent does not stop during the event) and thus the velocity of the agent is maximal during that event. For regressive optic flow (negative angular velocities), the average velocity during each event is less than maximum and for extreme angular velocities, it needs to stop for shorter durations to avoid collisions.
In order to quantitatively analyze how using regressive motion as a collision cue benefits agents to gain more fitness, we traced this particular agent's evolutionary line of descent (LOD) by following its lineage backward for 20,000 generations mutation by mutation until we reached the random agent that we used to see the initial population (see~\citealt{lenski2003evolutionary} for more details on how to construct evolutionary lines of descent for digitals). Figure~\ref{fig:lod_9} shows the fitness and the RCC value vs. generation for this agent's LOD. It is evident from these results that evolving this strategy benefits agents in gaining fitness compared to the rest of the population in this environment as high peaks of fitness occur at high RCC values and conversely, the fitness drops as the RCC value decreases.
Nevertheless, this strategy does not evolve all the time. Figure~\ref{fig:lod_p2} shows the fitness and RCC for all 20 replicates in the environment with penalty-reward ratio of 2. We can see that the mean fitness of all 20 replicates is around 20\% less than the fitness of the agent that evolved the  RCC strategy. The mean RCC value for all 20 replicates is also ~20\% less than that of an agent that evolved the RCC strategy.
The difficulty to evolve the RCC strategy is not limited to the number of runs in which this behavior evolved out of all replicates in some environment (we also tried running the experiment for longer evolutionary times but the results do not change significantly). Environmental conditions also play a key role in the evolution of this behavior. Figure~\ref{fig:rcc-env} shows the RCC value distribution for 20 replicates in five different environments. In order to calculate the RCC value in each replicate, we took the average of the RCC value in the last 1,000 generations on the line of descent to compensate for fluctuations. We observe that the RCC strategy only evolves in a narrow range of penalty-reward ratio, namely for PR=2 and PR=3. According to figure~\ref{fig:rcc-env}, higher values of penalty on the one hand discourage the agents from walking in the environment (they simply choose to remain stationary), and therefore prevent them from exploring the fitness landscape. Lower values for the penalty, on the other hand, result in indifference to collisions and thus, the optimal strategy (probably the local optimum) in these environments is to keep walking and ignore all collisions. In lower values of penalty, the RCC value is ~55\% which means they evolve to stop in obvious cases that end up in collision (if they keep moving without stopping at all, the RCC value should be 50). 
\section{Discussion}
We used an agent-based model of flies equipped with Markov brain networks that evolve via a GA to study the selective pressures and environmental conditions that can lead to the evolution of collision avoidance strategies based on visual information. We specifically tested cognitive models that invoke ``regressive motion saliency" and ``regressive motion as a cue for collision" to understand how flies avoid colliding with each other in two-dimensional walks. We showed that it is possible to configure the experiment in such a manner that ``regressive-collision-cue" (RCC) evolves as a strategy to avoid collisions. However, the conditions under which the RCC strategy evolved in our experiments are limited: the strategy only evolved in a narrow range of environmental conditions and even in those environments, it does not evolve all the time. In addition, we showed that from general principles, only a small percentage of events in which an agent perceives regressive optical flow eventually leads to a collision, so that RCC as sole strategy is expected to have a large false positive rate, leading to unnecessary stops. 

As discussed in the methods section, our experimental implementation is biased in such a way that {\em all} regressive motion events lead to a collision if the agent does not stop during that event. If the moving object's velocity direction is distributed uniformly randomly in all directions, the probability that a regressive event ends up in a collision is rather low ($\approx20\%$ in our implementations). Because the false positive rate of using regressive optical flow only as a predictor of collisions is liable to thwart the evolution of an RCC strategy, we biased our setup in such a way that the false-positive rate is zero. 
We now argue that this bias in our implementations does not significantly influence the outcome of our experiments. Consider an environment in which only a percentage of events with regressive motion end up in collision. This is similar to an environment with a lower penalty for collisions (as long as the strategy evolves at all) since the agent's fitness is scored at the end of its lifetime (all 100 events) not during each event.

However, there is a difference a between lower percentage of collisions in regressive events and lower penalty for collisions, namely a lower probability of collision in regressive motion events is equivalent to a higher amount of noise in the cue that the agent takes from the environment, compared to the case of lower penalties for collision. In other words, if 100\% of all regressive motion events lead to collisions, the agent associates regressive motion events with collisions with certainty. Thus, implementing the experiments with 100\% collisions in regressive motion events is tantamount to eliminating the noise in sensory information, which generally aids evolution. Compensating for noise in sensory information could also be achieved if we scored agents in every single event, and informed them about their performance in that event (feedback learning). We did not use feedback learning in the present experiments, but Markov gates equipped with such abilities exist, and we plan to use them in future experiments. 

We conclude that the evolution of ``regressive motion saliency" is unlikely to have happened only due to collision avoidance as the selective pressure. It is important to remember that walking is not the most frequent activity in fruit flies and furthermore that flies do not usually live in high density colonies and hence, do not find themselves on collision courses very often. It may be the case that components of this strategy (namely categorizing the optic flow as regressive or progressive) have evolved under different selective pressures entirely unrelated to the present test situation, and was further evolved to enhance collision avoidance with conspecifics while moving (a type of exaptation). For example, detecting predators is a strong selective pressure in the evolution of visual motion detection, including the categorization of that cue so as to take appropriate actions. It may be interesting to study the behavior of flies interacting with animals or objects that are not perceived as conspecifics.

This research has been supported in part by the National Science Foundation (NSF) BEACON Center under Cooperative Agreement DBI-0939454. We gratefully acknowledge the support of the Michigan State University High Performance Computing Center and the Institute for Cyber-Enabled Research (iCER).


\end{document}